\makeatletter
\newcommand*{\rom}[1]{\expandafter\@slowromancap\romannumeral #1@}
\makeatother
\documentclass[psf,proceedingpaper,accept,pdftex,oneauthor]{Definitions/mdpi} 
\firstpage{1} 
\pubvolume{7}
\issuenum{1}
\articlenumber{0}
\pubyear{2023}
\copyrightyear{2023}
\externaleditor{Academic Editor:Gonzalo Olmo 
 }
\datepublished{17 February 2023} 
\hreflink{https://doi.org/10.3390/ECU2023-14048} % If needed use \linebreak
\Title{On Cosmological Inflation in Palatini {$F(R,\phi)$} Gravity $^{\dagger}$}

\TitleCitation{On Cosmological Inflation in Palatini {$F(R,\phi)$} Gravity}

\Author{\hl{Mahmoud AlHallak} %MDPI: Please carefully check the accuracy of names and affiliations.
 \orcidA{}}

\AuthorNames{Mahmoud AlHallak}

\AuthorCitation{AlHallak, M.}
\address[1]{%
Physics Department, Faculty of Sciences, Damascus University, Damascus P.O. Box 30621, Syria; mahmoud.halag@unitedschool.ae}

\firstnote{\hangafter=1 \hangindent=1.05em \hspace{-0.82em}Presented at the 2nd Electronic Conference on Universe, 16 February–2 March 2023; Available online: \url{https://ecu2023.sciforum.net/}.}
\abstract{Single field inflationary models are investigated within Palatini quadratic gravity, represented by $R+\alpha R^2$,  along with a non-minimal coupling of the form $f(\phi) R$ between the inflaton field $\phi$ and the gravity. The treatment is performed in the Einstein frame, where the minimal coupling to gravity is recovered through conformal transformation. We consider various limits of the model with different inflationary scenarios characterized as  canonical slow-roll inflation in the limit $\alpha \dot{\phi}^2\ll (1+f(\phi)) $, constant-roll k-inflation for $\alpha \ll 1$, and slow-roll K-inflation for$ \alpha \gg 1$. A cosine and exponential potential are examined with the limits mentioned above and different well-motivated non-minimal couplings to gravity. We compare the theoretical results, exemplified by the tensor-to-scalar $r$ ratio and spectral index $n_s$, with the recent observational results of Planck 2018  and BICEP/Keck. Furthermore,  we include the results of a new study forecast precision with which $n_s$ and  $r$ can be constrained by currently envisaged observations, including CMB (Simons Observatory, CMB-S4, and~LiteBIRD).}

\keyword{Palatini {$F(R,\phi)$} gravity; constant-roll inflation; k-inflation; slow-roll inflation} 

\begin{document}
\section{Introduction}

\textls[-15]{The inflationary paradigm~\cite{Guth:1980zm} was suggested to solve several weaknesses of the standard Big Bang theory, including the horizon, flatness, and monopole problems. Furthermore, this paradigm produces the observed anisotropy of the cosmic \mbox{microwave~background.}}

As gravitational force dominates the universe during the inflationary period, general relativity and its alternatives provide frameworks for studying inflationary models. A popular alternative to general relativity is $F (R)$ gravity~\cite{Sotiriou:2008rp}, which introduces non-linear terms into the Ricci scalar $R$ to investigate different cosmological aspects such as \cite{Nojiri:2010wj,Nojiri:2017ncd,Myrzakulov:2015qaa} . 

This work aims to study single-field cosmological inflation within the quadratic form of $F(R)$ gravity. Due to the possibility of the well-motivated non-minimal coupling NMC between the inflaton field and gravity, we consider the generalized case of $F(R, \phi)$ gravity~\cite{Enckell:2018hmo}. Moreover, we perform the study from the perspective of Palatini's formulation of gravity, in which the metric and the affine connection are considered independent~variables. 

The paper is organized as follows. Section \ref{sec2} overviews the field equations within Palatini's $F(R, \phi) \equiv R+ \alpha R^2 + f(\phi) R$  gravity. In Section \ref{sec3}, the single field inflation is considered in three various limits of the model:  $2\alpha X\ll (1+f(\phi))$, $\alpha \gg 1$, and $\alpha \ll 1$; then, the cosine and exponential potentials with different well-motivated non-minimal couplings to gravity are presented as a case study.  Section \ref{sec4} is devoted to studying the observational results of considered models and comparing them to those of Planck 2018 (TT, EE, TE), BK15, and other experiments (lowE, lensing)~\cite{BICEP:2021xfz} in addition to a new study about the expected values of the observables according to experiments including CMB (Simons Observatory, CMB-S4, and LiteBIRD). Finally, we end up with a summary and discussions in Section \ref{sec5}.
%%%%%%%%%%%%%%%%%%%%%%%%%%%%%%%%%%%%%%%%%%
\section{Overview of {\boldmath$F(R,\phi)$} Gravity within Palatini Formalism}\label{sec2}
We begin with a general action of single-field inflation within $F(R, \phi)$ gravity as,
\begin{equation}
S=\int d^4x \sqrt{-g} \frac{1}{2}\Big\{ F(R,\phi)-g^{\mu\nu} \partial_\mu \phi \partial_\nu \phi - 2 V(\phi) \Big\}
\end{equation}

Here, $g$ is the determinant of the metric $g_{\mu\nu}$, $R = g^{\mu\nu}R_{\mu\nu}$ is the Ricci scalar, and $V(\phi)$ is the potential of the scalar field $\phi$. The function $F(R,\phi)$ is given as, 
\begin{equation}
F(R,\phi) \equiv \big(1+f(\phi)\big) R + \alpha R^2
\end{equation}
where $f(\phi)$ represents the non-minimal coupling to gravity. Several well-motivated forms of this function will be examined in this paper. The study will be performed in Palatini's formalism, where the Ricci tensor $R_{\mu\nu}(\Gamma,\partial \Gamma)$ is constructed using the connection $\Gamma$ which would be independent of the metric $g_{\mu\nu}$.  

In the following, we shall write the equivalent form of  $F(R,\phi)$ gravity to Brans--Dicke’s theory by introducing an auxiliary scalar field $\chi$ in such a way that the action is written as, 
\begin{equation}
\label{eq:S_action}
S=\int d^4x\sqrt{-g}\bigg\{\frac{1}{2}(\chi+f(\phi))R-\frac{(\chi-1)^2}{8\alpha}-\frac{1}{2} g^{\mu\nu}\partial_\mu \phi \partial_\nu \phi - V(\phi)\bigg\}
\end{equation}

The field $\chi$ appears with a kinetic coefficient of zero in the last action, meaning it has no dynamics.

We can utilize the above action form in the so-called Jordan frame, where the fields are non-minimally coupled to gravity; however, in this study, we switch to the Einstein frame, in which the action appears with minimal coupling; therefore, we can apply standard GR equations and inflationary solutions. 

In order to switch to the Einstein frame, we use the conformal transformations defined~as, 
\begin{equation}
\label{eq:conf_trans}
\tilde{g}_{\mu\nu}=\Omega^2(\phi,\chi)g_{\mu\nu}
\end{equation}
where $\Omega^2=(f(\phi) + \chi)$, then
the action, after dropping the tilde on $g_{\mu\nu}$ thereafter, becomes
\begin{equation}
\label{eq:sActionWchi}
S=\int d^4x \sqrt{-g}\bigg\{\frac{1}{2}R-\frac{1}{2} \frac{1}{\chi + f(\phi)}g^{\mu\nu}\partial_\mu \phi \partial_\nu \phi - \tilde{V}\bigg\}
\end{equation}
where the potential becomes:
\begin{equation}
\label{Vtilde}
\tilde{V}=\frac{8 \alpha V(\phi)+(\chi-1)^2}{8\alpha (f(\phi) + \chi)^2}
\end{equation}

Taking the variation of action (\ref{eq:sActionWchi}) with respect to the auxiliary field $\chi$, we can find a direct relationship between $\chi$ and the matter represented by $\phi$ as, 
\begin{equation}
\label{chi_phi}
\chi=\frac{1-\xi \phi^2 +8\alpha V + 2 \alpha \xi \phi^2  \partial^\mu \phi \partial_\mu \phi}{1-\xi \phi^2 - 2 \alpha \partial^\mu \phi \partial_\mu \phi}
\end{equation}

Then by substituting Equation (\ref{chi_phi}) into action (\ref{eq:sActionWchi}), we can find the final action in Einstein's frame as, 
\begin{equation}
S=S_{H-E}+S_\phi
\end{equation}
where $S_{H-E}$ is the standard Hilbert--Einstein gravitational action evaluated in the Einstein frame, and $S_\phi$ is the effective action of the scalar field, which is given as:
\begin{equation}
\label{final_action_phi}
S_\phi=\int d^4x \sqrt{-g} \Big\{ \sum_{j=1}^{j=2} G_j(\phi) X^j + \frac{V}{M_j(\phi)V^{j-1}}\Big\}
\end{equation}
with: 
\begingroup\makeatletter\def\f@size{9}\check@mathfonts
\def\maketag@@@#1{\hbox{\m@th\fontsize{10}{10}\selectfont\normalfont#1}}%  (或者把\fontsize{10}{10}\selectfont换成 \large)
\begin{equation}
\label{}
G_j(\phi)= \frac{(2 \alpha)^{j-1} (1+f(\phi))^{2-j}}{(1+f(\phi))^2+(8 \alpha V)},\quad M_j=(1+f(\phi))^{4-2j}+(8\alpha)^{j-1},\quad X=\frac{1}{2}g^{\mu\nu}\partial_\mu \phi \partial_\nu \phi 
\end{equation}\endgroup

The action \eqref{final_action_phi} shows that the contribution of the $R^2$ term turned the model into non-canonical scalar field model with kinetic term of the form $G(X,\phi)=\sum_{j=1}^{j=2} G_j(\phi) X^j $ and  effective potential $W=\sum_{j=1}^{j=2}\frac{V}{M_j(\phi)V^{j-1}}$. 

In the following sections, we will investigate different limits of the model with various inflationary scenarios. 
%%%%%%%%%%%%%%%%%%%%%%%%%%%%%%%%%%%%%%%%%%
\section{Single-Field Inflation}\label{sec3}
\subsection{Various Limitations of Action 9}
As part of this section, we will examine various limits of action (\ref{final_action_phi}) with regard to single-field inflationary models in  $F(R, \phi)$ gravity: 
\begin{itemize}
\item \paragraph{Limit \rom{1} : $2\alpha X\ll (1+f(\phi))$} \label{limit1}
This limit corresponds to the slow-roll inflationary paradigm with a canonical scalar field $\phi$. The action represents the model in this limit is given by:  
\begin{equation}
\label{}
S_{lim \rom{1}} = \int d^4x \sqrt{-g}\frac{1}{2}\bigg\{R - g^{\mu \nu} \partial_\mu \varphi \partial_\nu \varphi - 2 V_{eff}(\varphi(\phi))\bigg\}
\end{equation}
where $\varphi$ is re-scaled field defined as: 
\begin{equation}
\label{varphi}
 \bigg(\frac{d\phi}{d\varphi}\bigg)^2 = \sum_{j=1}^{j=2}(8\alpha)^{j-1}\big(1+f(\phi)\big)^{3-2j}V(\phi)^{j-1}
\end{equation}
The slow-roll parameters $\varepsilon$ and $\eta$, in addition to the observables $r$ and $n_s$, are summarized in Table \ref{tab1}. 

\item \paragraph{Limit \rom{2} : $\alpha \gg 1$}
Considering this limit, the model appears as a K-inflationary scenario. the action representing this situation is given by: 
\begin{equation}
\label{}
S_{lim \rom{2}} = \int d^4x \sqrt{-g}\frac{1}{2}\bigg\{R + \frac{1}{2} (g^{\mu \nu} \partial_\mu \chi \partial_\nu \chi)^2 - 2 V_{eff}(\chi(\phi))\bigg\}
\end{equation}
Now the re-scaled field $\chi$ is given as: 
\begin{equation}
\label{}
 \bigg(\frac{d\phi}{d\chi}\bigg)^4 = \frac{1}{2\alpha}\bigg[\big(1+f(\phi)\big)^2+8 \alpha V(\phi)\bigg]
\end{equation}
Table \ref{tab1} summarizes the model's main characteristics and observable formulas of this~limit. 
\item \paragraph{Limit \rom{3}: $\alpha \ll 1$}
Under the above limit, the model has the following action:
\begin{equation}
\label{}
S_{lim \rom{3}} = \int d^4x \sqrt{-g}\frac{1}{2}\bigg\{R +\alpha  (g^{\mu \nu} \partial_\mu \varphi \partial_\nu \varphi)^2- g^{\mu \nu} \partial_\mu \varphi \partial_\nu \varphi - 2 V_{eff}(\chi(\phi))\bigg\}
\end{equation}
where the re-scaled $\varphi$ is given by the Formula \eqref{varphi}.
Additionally, we treat the model at this limit according to the constant-roll scenario.

Again, the primary model's characteristics are summarized in Table \ref{tab1}. 
\end{itemize}
%\begingroup
%\setlength{\tabcolsep}{6pt} % Default value: 6pt
\renewcommand{\arraystretch}{1.3} % Default value: 1
\begin{table}[H] 
\caption{Main characteristics of the inflationary scenarios under study.\label{tab1}}
\newcolumntype{C}{>{\centering\arraybackslash}X}

\begin{adjustwidth}{-\extralength}{0cm}
%\centering %% If there is a figure in wide page, please release command \centering
\begin{tabularx}{\fulllength}{CcCC}
\toprule
\textbf{Inflationary Scenario}&\textbf{Slow-Roll Canonical Inflation} & \textbf{Slow-Roll K-Inflation}& \textbf{Constant-Roll K-Inflation}\\
\midrule
\textbf{Model Characteristics}&\textbf{Limit 1} & \textbf{Limit 2}& \textbf{Limit 3}\\
\textbf{}&\boldmath{$2\alpha X\ll (1+f(\phi))$} & \boldmath{$\alpha\gg1$}& \boldmath{$\alpha\ll1$}\\
\midrule
$3 H^2$		& $\frac{1}{2} \dot{\varphi}^2+ V_{eff}(\varphi)$			& $\frac{3}{4}\Dot{\chi}^4 + V_{eff}(\chi)$ & $\frac{1}{2}\Dot{\varphi}^2+\frac{3}{2}\alpha \Dot{\varphi}^4+ V_{eff}(\varphi)$\\
$\varepsilon$		& $\frac{1}{2} \frac{V^{\prime }_{eff}(\varphi)^2}{V_{eff}(\varphi)^2}$			& $\frac{1}{2} 3^{\frac{1}{3}}\frac{V^\prime_{eff}(\chi)^{\frac{4}{3}}}{V_{eff}(\chi)^{\frac{5}{3}}}$ & $-\frac{3\Dot{\varphi}^2}{4V_{eff}(\varphi)}(1+\alpha \Dot{\varphi}^2)$ \\
$\eta$	& $\frac{V^{\prime \prime}_{eff}(\varphi)}{V_{eff}(\varphi)}$			& $3^{\frac{1}{3}}\frac{{V}^{\prime \prime}_{eff}(\chi)}{(V_{eff}.V^\prime_{eff})^{\frac{2}{3}}}$ & $\frac{6 \alpha \beta\Dot{\varphi}^2}{1+6 \alpha \Dot{\varphi}^2}$\\
$n_s$	& $1 - 6 \varepsilon + 2 \eta$			& $1-\frac{1}{3} (4\eta - 16 \varepsilon)$ & $1+2\frac{2\varepsilon-\beta-\eta}{1+\varepsilon}$\\
$r$	& $16 \varepsilon$			& $\frac{16}{3} \varepsilon$ & $E C_A^{3+\beta} \bigg(\frac{(1+6 \alpha \Dot{\varphi}^2)\Dot{\varphi}^2}{W(\varphi)}\bigg)$\\
\bottomrule
\end{tabularx}
\end{adjustwidth}
\noindent{\footnotesize{\textsuperscript{\hl{1} %MDPI: There is not superscript 1 in the table body, please revise.
} In the case of contact-roll k-inflation, $\beta=\frac{\ddot{\varphi}}{H \Dot{\varphi}}=constant$, $E=\bigg(\frac{\sqrt{6}\Gamma(\frac{3}{2})}{\Gamma(\frac{3}{2}+\beta)2^\beta}\bigg)^2,
$ and $C_A=\frac{1-2\Dot{\varphi}^2}{1-6\Dot{\varphi}^2}$.}}
\end{table}
%\endgroup
%%%%%%%%%%%%%%%%%%%%%%%%%%%%%%%%%%%%%%%%%%
\subsection{Case Study: Cosine {and} Exponential Potentials}
This section summarizes the three prominent cases examined in~\cite{AlHallak:2021hwb,AlHallak:2022gbv}, including cosine potential with periodic/non-periodic non-minimal coupling and the exponential potential with an inverse exponential coupling to gravity. A summary of the three study cases is presented in Table \ref{tab2}.

The inflationary model corresponds to the cosine potential of cases \rom{1} and \rom{2}, mainly known as Natural inflation~\cite{Freese:1990rb}. The inflaton field is modeled directly on the QCD axion, albeit with a different mass scale. 

Following~\cite{CHERNIKOV:1968}, we can justify the $\xi \phi^2 R$'s NMC term of case \rom{1} by considering that the corresponding Klein--Gordon equation must include such a term for a massless scalar field in curved spacetime with a specific value of $\xi=\frac{1}{6}$ in order to remain conformally symmetric. Considering the ``not yet computed'' loop quantum effects, depending on the ``unknown'' microscopic theory, pushes us to take $\xi$ as a free parameter. Furthermore, in~\cite{Reyimuaji:2020goi} CP conservation and simplifying arguments were presented to support the form of this NMC.  The NMC term of case \rom{2} is well motivated in~\cite{Ferreira:2018nav}, where the authors assumed that the NMC term respects the shift symmetry $(\phi \rightarrow \phi + 2 \pi f)$ of the axion potential.

 Referring to case \rom{3}, the authors of refs.~\cite{Chamoun:2006jj,AlHallak:2016wsa} clarified that temporal variation of the strong coupling constant, encoded in a non-canonical scalar field $\epsilon$, can generate an inflationary epoch. However, assuming a non-minimal coupling to the gravity of form $\xi \epsilon^2 R$, we can obtain the potential and canonical terms of case \rom{3} after re-scaling the field $\epsilon$ as $\epsilon=\exp{\ell \phi}$~\cite{AlHallak:2021sic}.	

\begingroup
\setlength{\tabcolsep}{6pt} % Default value: 6pt
\renewcommand{\arraystretch}{1.3} % Default value: 1
\begin{table}[H] 
\caption{The considered potentials, along with the corresponding non-minimal coupling functions.\label{tab2}}
\newcolumntype{C}{>{\centering\arraybackslash}X}
\begin{tabularx}{\textwidth}{CCCC}
\toprule
	&\textbf{Case \rom{1}}	& \textbf{Case \rom{2}}	& \textbf{Case \rom{3}}\\
\midrule
\textbf{\hl{Potential:} %MDPI: Please confirm if the bold should be retained.
 $V(\phi)=$}	&$V_0 \bigg(1+\cos(\frac{\phi}{f})\bigg)$		& $V_0 \bigg(1+\cos(\frac{\phi}{f})\bigg)$			& $V_0 \exp({-2\ell \phi})$\\
\textbf{\hl{NMC term:} $F(\phi)=$}	&$-\xi \phi^2$		& $\lambda(1+\cos{\phi/f)}$			& $-\xi \exp{(2\ell \phi)}$\\
\bottomrule
\end{tabularx}
\end{table}
\endgroup
%%%%%%%%%%%%%%%%%%%%%%%%%%%%%%%%%%%%%%%%%%%
\section{Results}\label{sec4}
We present the main findings of the study in this section. In our survey of inflationary models within a modified $F(R, \phi)$ gravity, we realized an interesting fact regarding quadratic and non-minimal coupling terms. The gravitational quadratic term $\alpha R^2$ influences the scalar--tensor ratio $r$, where increasing $\alpha$ values decrease $r$ values. Meanwhile, the non-minimal coupling $f(\phi)R$ has a major effect on $n_s$, where increasing the NMC parameter leads to an increase in $n_s$. These findings are summarized in Equation \eqref{model_finds}.
\begin{equation}\label{model_finds}
\alpha \nearrow \Longrightarrow r \searrow  \quad,\quad \xi \nearrow \Longrightarrow n_s \nearrow 
\end{equation}

Since the parameters $\alpha$ and $\xi / \lambda$ are pretty loose, the model can fit the higher accurate future observations of $r$ and $n_s$ since having both terms together serves as a ``focusing'' tool affecting the $r$ and $n_s$ values separately. In order to affirm the focusing effect, we add a forecasting study about the future results of CMB (Simons Observatory, CMB-S4, and LiteBIRD), optical/near infra-red (DESI and SPHEREx), and 21 cm intensity mapping (Tianlai and CHIME) surveys~\cite{Bahr-Kalus:2022prj}.

\begin{figure}[H]

\begin{adjustwidth}{-\extralength}{0cm}
\centering %% If there is a figure in wide page, please release command \centering
\includegraphics[width=15 cm]{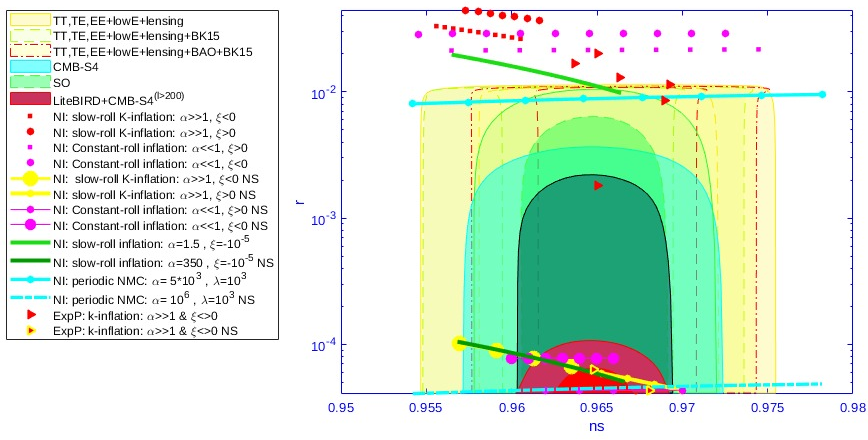}
\end{adjustwidth}
\caption{\hl{A plot} %MDPI: Please use minus sign instead of hyphen to mark negative number in image. For example, “-200” should be “−200”.
 of tensor-to-scalar ratio $r$ and spectral index $n_s$ for various cases of Table \ref{tab2}, within limits I, II, and III. the theoretical results of different cases and limitations are compared to the joint 68 and 95$\%$ CL regions for $n_s$, and $r$, obtained from Planck 2018 and lensing data alone, and their combinations with BICEP2/Keck Array (BK15) and BAO data, in addition to a  forecast posterior contours of the  CMB experiments (Simons Observatory, CMB-S4, and LiteBIRD) considered \hl{in} %MDPI: Please make sure that permission has been obtained and there is no copyright issue.
 \cite{Bahr-Kalus:2022prj}. NI, ExP, and NS stand for Natural inflation, Exponential Potential, and New Scan, respectively. }
\label{figure1}
\end{figure}

In Figure \ref{figure1}, we reproduce most of the results of~\cite{AlHallak:2021hwb,AlHallak:2022gbv} regarding the cases of \mbox{Table \ref{tab2}}. The exemplified results of Figure \ref{figure1} were largely in agreement with the observations of Planck 2018~\cite{AlHallak:2021hwb,AlHallak:2022gbv}. We see, however, that the opposite occurs when Planck 2018 and lensing data alone and their combinations with BICEP2/Keck Array (BK15) and BAO data are taken into account. A large majority of the results are excluded since they do not meet observational constraints. Although adding the results of the Ref. ~\cite{Bahr-Kalus:2022prj} makes the theoretical results further away from the observational ones, we do so to illustrate the model's strength and show its focusing ability. We performed a new scan on the parameters and successfully attained a coincidence with the observational results again, as we can see in  Figure \ref{figure1}. Table \ref{table3} clarifies the scanned ranges of parameters and the scopes of observables $n_s$ and $r$for each of the considered cases.

%\begingroup
%\setlength{\tabcolsep}{2.3pt} % Default value: 6pt
%\renewcommand{\arraystretch}{1.3} % Default value: 1
% \usepackage{multirow}
\begin{table}[H]
\caption{The scanned ranges of parameters and the scopes of observables $n_s$ and $r$ for each of the considered cases in Figure \ref{figure1}.}
\label{table3}

\begin{adjustwidth}{-\extralength}{0cm}
%\centering %% If there is a figure in wide page, please release command \centering
\begin{tabularx}{\fulllength}{CCCCCCC}
\toprule
\multicolumn{1}{c}{\textbf{Case}}                                                                       & \boldmath$\alpha$          & \boldmath$\xi / \lambda (\times 10^{-5})$ & \multicolumn{1}{c}{\boldmath$V_0$} & \boldmath$f/\ell$                          & \multicolumn{1}{c}{\boldmath$n_s(\times10^{-2})$} & \multicolumn{1}{c}{\boldmath$r(\times 10^{-3})$} \\ \midrule
\multirow{2}{*}{\begin{tabular}[c]{@{}l@{}}Limit I, Case I\\ Slow-roll NI\end{tabular}}        & $1.50$            & $-1.00 $                         & $0.24$                    & $7.00$                            & \hl{$[95.6-96.6]$} %MDPI: Please confirm if minus between numbers should be en dash. If so, please modify all in this table
                         & $[19.7-979]$                            \\
                                                                                               & $350$             & $-1.00$                          & $0.24$                    & $75.0$                            & $[95.6-96.6]$                         & $[0.105-0.05]$                          \\ \midrule 
\multirow{2}{*}{\begin{tabular}[c]{@{}l@{}}Limit I, Case II\\ Slow-roll NI\end{tabular}}       & $5\times 10^3$    & $1000$                           & $1000$                    & $1000$                            & $[97.8-95.4]$                         & $[9.53-8.03]$                           \\
                                                                                               & $1\times10^6$     & $1000$                           & $1000$                    & $1000$                            & $[97.8-95.4]$                         & $[0.049-0.041]$                         \\ \midrule 
\multirow{2}{*}{\begin{tabular}[c]{@{}l@{}}Limit II, Case I\\ K-inflation NI\end{tabular}}     & $250$             & $[90-1.5]$                       & $0.0002$                  & $45.0$                            & $[95.7-96.2]$                         & $[44.5-36.8]$                           \\
                                                                                               & $21000$           & $[4.9-5.2]$                      & $0.006$                   & $45.0$                            & $[96.5-97.0]$                         & $[0.059-0.043]$                         \\ \midrule 
\multirow{2}{*}{\begin{tabular}[c]{@{}l@{}}Limit II, Case III\\ K-inflation ExP\end{tabular}} & $500$             & \multicolumn{2}{c}{$z=0.505$}                                & \multicolumn{1}{l}{$[1.24-1.44]$} & $[97.1-96.4]$                         & $[1.68-61.1]$                           \\
                                                                                               & $30{,}000$           & \multicolumn{2}{c}{$z=0.505$}                                & \multicolumn{1}{l}{$[1.30-1.40]$} & $[96.8-96.5]$                         & $[0.043-0.064]$                         \\ \midrule 
\multirow{2}{*}{\begin{tabular}[c]{@{}l@{}}Limit III, Case I\\ Constant-roll NI\end{tabular}}  & $1\times10^{-5}$  & $10$                             & $11.0$                    & $9.00$                            & $[96.8-95.4]$                         & $[29.1-28.7]$                           \\
                                                                                               & $1\times 10^{-5}$ & $10$                             & $2.00$                    & $9.00$                            & $[96.6-96.0]$                         & $[0.078-0.077]$                         \\ \bottomrule
\end{tabularx}
\end{adjustwidth}
\end{table}
%\endgroup

\section{Discussion {and} Conclusions}\label{sec5}
In this work, we studied the single-field inflationary models within the palatini quadratic gravity with non-minimal coupling between the inflaton field and the gravitational background. Two specific inflationary models are considered and investigated in the study: natural inflation inspired by particle physics and exponential potential model inspired by VLS. In addition, the study assumed different well-motivated functions of non-minimal coupling to gravity. 

The study shows an interesting impact of the considered extended gravitational model on the inflationary scenario, where the gravitational model exhibits a focusing effect regarding the observable quantities exemplified as $n_s$ and $r$. We scanned the parameter spaces of the models in order to compare their results with the observational ones represented by Planck 2018 and BICEP2/Keck Array. To investigate the focusing effect of the model on a more profound level, we include the work with a prediction study about the future expected values of the observables, and we show that the gravitational model, by its focusing ability, succeeded in accommodating the new challenging constraints. 
%%%%%%%%%%%%%%%%%%%%%%%%%%%%%%%%%%%%%%%%%

\vspace{6pt} 

%%%%%%%%%%%%%%%%%%%%%%%%%%%%%%%%%%%%%%%%%%
%% optional
%\supplementary{The following supporting information can be downloaded at:  \linksupplementary{s1}, Figure S1: title; Table S1: title; Video S1: title.}


%%%%%%%%%%%%%%%%%%%%%%%%%%%%%%%%%%%%%%%%%%
\funding{This research received no external funding.}

\institutionalreview{\hl{ }.}%In this section, you should add the Institutional Review Board Statement and approval number, if relevant to your study. You might choose to exclude this statement if the study did not require ethical approval. Please note that the Editorial Office might ask you for further information. Please add “The study was conducted in accordance with the Declaration of Helsinki, and approved by the Institutional Review Board (or Ethics Committee) of NAME OF INSTITUTE (protocol code XXX and date of approval).” for studies involving humans. OR “The animal study protocol was approved by the Institutional Review Board (or Ethics Committee) of NAME OF INSTITUTE (protocol code XXX and date of approval).” for studies involving animals. OR “Ethical review and approval were waived for this study due to REASON (please provide a detailed justification).” OR “Not applicable” for studies not involving humans or animals.

\informedconsent{\hl{ }.}%Any research article describing a study involving humans should contain this statement. Please add ``Informed consent was obtained from all subjects involved in the study.'' OR ``Patient consent was waived due to REASON (please provide a detailed justification).'' OR ``Not applicable'' for studies not involving humans. You might also choose to exclude this statement if the study did not involve humans.%Written informed consent for publication must be obtained from participating patients who can be identified (including by the patients themselves). Please state ``Written informed consent has been obtained from the patient(s) to publish this paper'' if applicable.

\dataavailability{\hl{ }.}%In this section, please provide details regarding where data supporting reported results can be found, including links to publicly archived datasets analyzed or generated during the study. Please refer to suggested Data Availability Statements in section ``MDPI Research Data Policies'' at \url{https://www.mdpi.com/ethics}. If the study did not report any data, you might add ``Not applicable'' here. 

\acknowledgments{I am very grateful to N.Chamoun for his help and support.}

\conflictsofinterest{The author declares no conflicts of interest.} 

%%%%%%%%%%%%%%%%%%%%%%%%%%%%%%%%%%%%%%%%%%

\begin{adjustwidth}{-\extralength}{0cm}

\reftitle{References}


%\bibliography{your_external_BibTeX_file}

%=====================================
% References, variant B: internal bibliography
%=====================================

\PublishersNote{}
% If authors have biography, please use the format below
%\section*{Short Biography of Authors}
%\bio
%{\raisebox{-0.35cm}{\includegraphics[width=3.5cm,height=5.3cm,clip,keepaspectratio]{Definitions/author1.pdf}}}
%{\textbf{Firstname Lastname} Biography of first author}
%
%\bio
%{\raisebox{-0.35cm}{\includegraphics[width=3.5cm,height=5.3cm,clip,keepaspectratio]{Definitions/author2.jpg}}}
%{\textbf{Firstname Lastname} Biography of second author}

% For the MDPI journals use author-date citation, please follow the formatting guidelines on http://www.mdpi.com/authors/references
% To cite two works by the same author: \citeauthor{ref-journal-1a} (\citeyear{ref-journal-1a}, \citeyear{ref-journal-1b}). This produces: Whittaker (1967, 1975)
% To cite two works by the same author with specific pages: \citeauthor{ref-journal-3a} (\citeyear{ref-journal-3a}, p. 328; \citeyear{ref-journal-3b}, p.475). This produces: Wong (1999, p. 328; 2000, p. 475)

%%%%%%%%%%%%%%%%%%%%%%%%%%%%%%%%%%%%%%%%%%
%% for journal Sci
%\reviewreports{\\
%Reviewer 1 comments and authors’ response\\
%Reviewer 2 comments and authors’ response\\
%Reviewer 3 comments and authors’ response
%}
%%%%%%%%%%%%%%%%%%%%%%%%%%%%%%%%%%%%%%%%%%
%\PublishersNote{}
\end{adjustwidth}

\end{document}